\documentclass[twocolumn,showpacs,showkeys,preprintnumbers,amsmath,amssymb]{revtex4}
\usepackage{graphicx}
\usepackage{dcolumn}
\usepackage{bm}
\begin{document}
\title{\large Spontaneous Symmetry Breakdown and Critical Perspectives of Higgs Mechanism}
\author{Nils M. Bezares-Roder$^1${\footnote{e-mail: Nils.Bezares@uni-ulm.de}} and Hemwati Nandan$^2${\footnote{e-mail: hnandan@cts.iitkgp.ernet.in \\ \\ {** This article is dedicated to Professor Heinz Dehnen on his $70{th}$ birth anniversary  for his pioneering contribution to the Higgs field gravity. \\  \\ Published in the {\it Indian Journal of Physics} {\bf 82(1)} (2008) 69-93.}}}}
\vspace{0.2cm} \affiliation{$1.$ Institut f$\ddot{u}$r
Theoretische
Physik, Universit$\ddot{a}$t Ulm, $89069$ Ulm, Germany.  \\
$2.$ Centre for Theoretical Studies, Indian Institute of
Technology, $721 302$ Kharagpur, India.}
\begin{abstract}
\noindent The foundations of the mass generation mechanism of
particles are reviewed. The Spontaneous Symmetry Breaking (SSB)
process within the standard model (SM) and the minimal
supersymmetric standard model (MSSM) is used to explore the present
status of the Higgs Mechanism along with the constraints in
detecting the Higgs particles in experiments. The possible
explanations and generalizations for the case that the Higgs
particles should not appear or to couple the Higgs Mechanism
(because of their gravitational nature of interaction) are also
discussed in detail in view of the Higgs field gravity.
\end{abstract}
\vspace{0.2cm}  \maketitle
\section{Introduction}
\noindent The mass of any physical object is thought to be an
entity which quantifies its amount of matter and energy. However,
conceptually, it is defined as inertial and gravitational (passive
and active) mass. The inertial mass is indeed a measure of an
object's resistance to the change of its position due to an
applied force. On the other hand, the passive gravitational mass
measures the strength of an object's interaction with the
gravitational field, while the active one is a measure of the
strength of the gravitational field due to a particular object.
However, Einstein's principle of equivalence asserts the equality
of the gravitational $(m_g)$ and inertial mass $(m_i)$ of a body.
In fact, by now all the experiments have failed to make any
difference between them \cite{Weinberg, Hartlee}. More generally,
such equivalence between the inertial and gravitational mass is
included, for instance, in the weak equivalence principle (WEP),
which confirms the universality of free fall such that all bodies
in a given gravitational field and at the same space-time point
would undergo the same acceleration. Further, the strong
equivalence principle (SEP) is a generalization in the sense that
it governs all the effects of the gravitational interaction on all
physical systems and holds for all the \emph{laws of nature}. In
fact, the WEP replaces the \emph{laws of nature} (which might be
the case for SEP) by the \emph{laws of motion of freely falling
bodies} \cite{Weinberg}. Independently of the equivalence of
masses, however, the origin of mass itself is another conceptual
problem of physics to solve, and the mechanism through which
particles acquire mass is indeed an important subject from the
point of view of the basic constituents and interactions among
them in nature. In this context, the mass generation is identified
with the symmetry breakdown of the Lagrangian corresponding to a
particular theory where the mass comes as a consequence of the
symmetry loss and features of the self-interactions in the
standard model (SM), which is perhaps the most celebrated theory
of modern elementary particle physics. The SM is a full
relativistic quantum field theory, and it has been indeed incredibly
successful in describing the electromagnetic, weak and strong
interactions between the basic constituents (quarks and leptons)
with the symmetry group $ G_{SM} \equiv SU(3)_{C} \otimes
SU(2)_{W} \otimes U(1)_{Y}$ down to the distances as
small as $10^{-16} cm$. In SM, the interaction of the
constituents of matter with the Higgs field allows all the
particles to have different mass \cite{Kan93}. The mass of the
particles obtained via the so-called Higgs Mechanism is then
proportional to the vacuum expectation value (VEV) of the Higgs
field so that mass can be given in terms of the parameters of the
Higgs potential. However, in GR it is still not possible to
explain the origin of mass from the curved space-time, and the mass
is used as a parameter with the equivalence $m_g \equiv m_i \equiv M$. The SM, therefore, provides a unique way to explain the acquisition of
mass by basic constituents of matter via Higgs Mechanism
\cite{Peskin}. Moreover, the cosmological consequences of mass
production still demands further explanation and a unified theory
in this context is still lacking. However, some physical aspects
to completely solve the issue within gauge-gravitation
theories, Supersymmetry (SuSy), Supergravity (SuGra) and Loop
Quantum Gravity (LQG) are really promising. It is noteworthy that
the appearance of mass in SM by the virtue of Higgs Mechanism
comes in a natural way as well as it co-exists peacefully in
various known processes of the physical world as predicted by SM
itself. Unfortunately, some of the basic aspects of the Higgs
Mechanism are still unknown as the Higgs particle is still not an
experimental reality. However, with the developments in SuSy
(which emphasize a symmetry between fermions and bosons), it leads
to the possibility to cancel unphysical quadratic divergences in
the theory as well as it provides an answer to the hierarchy
problem between the electro-weak (EW) ($\sim 10^2$GeV) and Planck ($\sim 10^{19}$GeV) scale  \cite{Kan93}. Therefore, the
supersymmetric version of the SM may play an important role to
stabilize the hierarchy against quantum corrections, and in the
minimal supersymmetric SM (MSSM) with the radiative EW symmetry
breaking, the stability of the Higgs field leads to mass
generation to be around the EW scale. Remarkably enough, the
problem of mass generation and its explanation is still a very
important subject which needs to be explored in view of the
various developments in modern physics, and it is certainly
not a closed chapter for further discussions. In fact, the Higgs
Mechanism along with the search of the Higgs particles at higher
and higher energies has narrowed down the scope for other theories
in this regard and has become a natural tendency to find an
appropriate answer to understand the mass generation \cite{Kan93}.\\
In the present article, the problems associated with the mass
generation are revisited from the perspectives of the different
well known mass-containing Lagrangians. The Higgs Mechanism in
view of the SM is summarized with the current experimental status.
The various phenomenological aspects related to the Higgs
Mechanism (in view of the different unification schemes of the
fundamental interactions) are reviewed. The gravitational-like
interactions and the possibility without interacting Higgs
particles, which puts some constraints on Higgs Mechanism, are
also discussed by the virtue of the Higgs field gravity. The impact of the Higgs scenario on the physical world is concluded along with its
possible future prospects.
\section{The mass generation and different symmetry breaking modes}
\noindent In order to have a discussion on the mass generation
mechanism within the notions of the analytical mechanics, let us
first write Hamilton's principle of the stationarity (or least)
action in the following form,
\begin{equation}
\delta \int \,  {\cal L} \, \sqrt{-g} \, d^4x \, \equiv \, 0.
\end{equation}
In fact, there are two possible ways to introduce the mass term in a
particular theory: \\
 \noindent (i) With an additional term $({\cal
L}_m)$ containing the mass in the general Lagrangian ${\cal L}$.\\
\noindent(ii) With the SSB via an extra term ${\cal L}_H \equiv
{\cal L}_5$ having a $5^{th}$ symmetry-breaking force.
\\
\noindent It is possible to achieve the well-known equations (viz
Schr$\ddot{o}$dinger, Klein-Gordon and Dirac equation) in
non-relativistic as well as in relativistic quantum mechanics (RQM)
with the aforesaid first choice. The Lagrangians from which these
equations can be derived, actually contain the mass terms and have
the following form in the natural system of units,
\begin{equation}
{\cal L}_{SE}= -\frac{1}{2M}\, \psi^*_{,k} \psi_{,k}
+\frac{i}{2}(\psi^*\, \psi_{,t}-  \psi \, \psi^*_{,t})- V(\psi \psi^*)
,\label{Sch}
\end{equation}
\begin{equation}
{\cal L}_{KG}= \frac{1}{2}\,  \left(\psi^*_{,\mu} \psi^{,\mu}- M^2
\psi^* \psi \right),\label{KG}
\end{equation}
\begin{equation}
 {\cal L}_{D}= \frac{i}{2} \,( \bar{\psi} \gamma^\mu\,
\psi_{,\mu}- \, \bar{\psi}_{,\mu} \gamma^\mu\,
\psi )\, - M \, \bar{\psi}\psi, \label{Dir}
\end{equation}
where $M$ denotes the mass and ${,\mu}\equiv \partial _\mu $, while
$\gamma^\mu$ represent the well-known Dirac matrices. The problem with this choice, however, relies on the well tested fact of parity
violation (viz the CP violation in the $\beta$-decay in Wu-like
experiments in the electro-weak interactions). This violation cannot be achieved by adding mass by hand because in the equations
(\ref{Sch}-\ref{Dir}), the left- and right-handed particles couple
all in the same way to vector-bosons in order to preserve the
gauge invariance \cite{Kan93}. Moreover, a massive propagator
(which gives the probability amplitude for a particle to travel
from one place to another in a given time, or to travel with a
certain energy and momentum, in this case for massive virtual
particles) does not lose its longitudinal term and as well does not
transform in a (transversal), massless, one in the limit $M
\rightarrow 0$. As a consequence of this, most closed Feynman
graphs diverge, which makes the theory non-renormalizable. It is
therefore needed to have a theory with the requirements of
renormalizablity and which can be achieved by the spontaneous
symmetry breakdown, for which the existence of an extra scalar field
(the Higgs field) is needed to make the theory (e.g. SM)
mathematically consistent \cite{Kan93,Vel77}. The characteristics
of different symmetry-breaking modes may therefore be defined from
the point of view of the parity violation and renormalization. For
the parity violation and well behaved propagators, the best choice
is the symmetry breaking where the mass is produced as a
consequence of the \emph{loss} of symmetry and self-interactions.
The requirement for such breakdown of symmetry also demands
another gauge invariant (i.e. current-conserving) term in the
Lagrangian, which is identified with the interaction of the
spontaneous production of mass. The symmetry breakdown can be
given through different modes, depending on the properties of the
ground state. In different quantum field theories, the ground
state is the vacuum state, and it is therefore important to check
the response of the vacuum state to the symmetry breaking. As
such, there are three main modes \cite{Gui91} as given below: \\
 \,\, \, (i) The \emph{Wigner}-Weyl mode,
\\ \vspace{0.08cm}
(ii)  The Nambu-\emph{Goldstone} mode,\\
 \vspace{0.08cm}
 (iii) The Higgs mode.\vspace{0.2cm}
\\In these processes, the symmetry
group $G$ breaks down to a rest-symmetry group $\tilde{G}$ (i.e.
$G\rightarrow\tilde{G})$ with $\tilde{G}=\bigcap_{r=1}^n
\tilde{G}_r$ where $n>1$ is valid for more than one breaking
process. For instance, in the SM, the breaking $SU(3)_C \otimes SU(2)_W\otimes U(1)_Y \rightarrow SU(3)_C \otimes U(1)_{em}$ is valid,  while for the grand unified theory (GUT) under $SU(5)$, the breaking
$SU(5)\rightarrow SU(3)_C\otimes SU(2)_W\otimes U(1)_Y$ also takes place at about $10^{15} GeV$. However, another interesting example of symmetry-breaking comes from the fundamental asymmetry between space and time which is found in the signature of the relativistic metric \cite{Wet05} and is mostly given in \emph{ad hoc} manner. Such asymmetry may be generated as a property of the ground state following a symmetry breakdown in the universe from which the structure of the quantum field theory and gravitational field equations would be derivable.\\
In particular, the \emph{Wigner}-Weyl mode is the most usual
symmetry-breaking mode in quantum mechanics (QM), with a real
invariant vacuum which can be identified with the classical one as
follows,
\begin{equation}
U|0> \, = \, |0>.
\end{equation}
The \emph{Wigner}-Weyl mode is indeed related to the existence of
degeneracy among particles in the multiplets and the violation of
which enforces an explicit symmetry breakdown in the Hamiltonian
$H$. Such situation appears in the Zeeman effect where turning-on of
the external fields causes the breakdown of the rotational symmetry.
One more example of the \emph{Wigner}-Weyl mechanism may be seen in
the breaking of $SU(3)_C$ to $SU(2)_{W}$ due to the effect of
hypercharges, and which further breaks in $U(1)$ because of Coulomb
interaction. However, the $U(1)$-symmetry remains unbroken because
of the current-conservation law \cite{Gui91}. Further, in both the
Nambu-\emph{Goldstone} and Higgs modes, the symmetry is actually not
lost but camouflaged and hidden in the background of the mass
generation. However, these two modes differ from each other through
their gauge-symmetry while both of them are given by the vacuum
defined as follows,
\begin{equation}
U|0> \, \neq \, |0>.
\end{equation}
It is worth mentioning that the Nambu-\emph{Goldstone} mode works
\emph{globally} while the Higgs mode acts \emph{locally} in view
of gauge invariance. The main difference between them is that in
the Nambu-\emph{Goldstone} Mechanism both massive (Higgs) and
massless (Goldstone) particles (generally bosons) appear, while in
the Higgs Mechanism only the massive particles are present and the
mass acquisition of gauge bosons is at the cost of the Goldstone
particles, which are gauged away unitarily. The degrees of freedom
of the massless particles, however, do not disappear from the
physical spectrum of the theory. In general sense, the gauge
fields absorb the Goldstone bosons and become massive while the
Goldstone bosons themselves become the third state of polarization
for massive vector-bosons. This interpretation is analogous to the
\emph{Gupta-Bleuler} Mechanism where, with the quantization of a
massless field $A^\mu$, the temporal degree of freedom of $A^\mu$
(i.e. $A^0$) cancels with longitudinal space-like components of
$p_\mu \cdot {\bf A}$ in a way that $A_\mu$ becomes the
transversal components of ${\bf p} \times {\bf A}$ \cite{Gui91}.
In the Higgs Mechanism, the Goldstone mode cancels the time-like
components of the gauge fields in such a way that the three
space-like components remain intact and $A_\mu$ behaves like a
massive vector-boson. The mass generation by this sort of way can
best be identified in the Meissner effect in conventional
superconductivity, and the SSB may therefore be applicable in
explaining such mechanism also in non-relativistic theories
\cite{Gre95}. However, an analogy between the Higgs Mechanism and
the Meissner effect may be explained in terms of the Yukawa-Wick
interpretation of the Higgs Mechanism where the Goldstone bosons
at unitary gauge vanish because of the existence of long-ranged
forces while their short-ranged behavior may be transcribed by
Yukawa's theory for the massive fields. The condensed
electron-pairs (the Cooper pairs) in the ground state of a
superconductor may then be identified with a Higgs field which
leads the magnetic flux expulsion with a finite range given by the
penetration depth, which is basically the reciprocal effective
mass acquired by the photons. For instance, in a Scalar-Tensor
Theory (STT) of gravitation with symmetry breaking, which is derivable
as the simplest Higgs-curvature coupled theory \cite{Hil87, Bij95}
and is based on the analogous properties of the Higgs and
gravitation \cite{Deh91, Bij94}, the Higgs field of the theory has
a finite range which is the inverse of its Higgs field mass
\cite{Bez07a}. There, analogies between the
Higgs field for the Schwarzschild metric and the London equations
for the Meissner effect also exist. In general, the coupling between the superconductor and Higgs is of more profound nature since it helps
in modern contributions to understand SM, especially in context of
dual QCD where the Higgs field with magnetic charge leads to
the Meissner effect of color electric flux which provides
a unique way to understand the quark confinement mechanism in the
background of the magnetic charge condensation \cite{Mand, Hn1}.
\section{The Higgs mechanism and unitary gauge}
\noindent The mass generation through an interaction with a
non-empty vacuum can be traced back to the $\sigma$-model proposed
by Schwinger where the $\sigma$ and $\varphi_i$ ($i=1, \ldots, 3$)
lead to the appearance of three massive and one massless vector
bosons. The $\sigma$-model seems typical in view of the physical
economy in comparison to the Higgs Mechanism, which demands the
appearance of only one scalar field $\phi$ \cite{Hig64}. In fact,
the scalar multiplet in the SM belongs to a doublet representation
of the gauge group in the following form,
\begin{equation}
\phi={{\phi ^{+}} \choose {\phi ^{0}}},
\end{equation}
which is defined with a non-trivial vacuum state having the
characteristics of symmetry breaking of the gauge group $G$ to the
rest-symmetry of the isotropy group $\tilde{G}$. The complex field
$\phi^0$ can be further re-written in terms of real fields (i.e.
$\phi^0=(\tilde{\sigma}+ i\chi) / \sqrt{2})$. With the
spontaneous breakdown of the gauge symmetry, the minimal value of
the energy-density $u$ is taken by the ground state value $\phi_0=v$
with $<\tilde{\sigma}>=v$. The $\tilde{\sigma}$ and $\chi$- fields
may then be identified with the Higgs particles and Goldstone bosons
respectively. The symmetry of the Lagrangian is then broken when
particles fall from their false vacuum (with $\phi=0$) to the real
one ($\phi=v$). In general, for such SSB, the less energy is
required to generate a new particle (i.e. Higgs particle) with the
associated features of the self-interaction. With Higgs bosons as
neutral particles, the photons are not able to \emph{see} them and
remain massless in electrodynamics, while, however, the neutral
$Z$-bosons couple to Higgs bosons via a Weinberg-mixture with
charged $W$-boson fields.\\
The Higgs mode, in fact, does not need to violate parity, while
this indeed occurs in the $\sigma$-model \cite{Kan93}.
Nevertheless, this violation is given in SM through the isospin
scalar field $\phi$ as a doublet in its iso-vectorial form
instead of only an iso-scalar with $\phi=vN$, where the unitary vector $N$ in the isospin-space satisfies $N^\dagger N=1$. On the other hand, the right-handed bosonic multiplets are only iso-scalar, while
left-handed ones are iso-doublets (for up and down states). The
multiplets acquire mass through the components of $N$, which in the
unification model (viz $SU(5)$ for instance) is matrix-valued for
the first symmetry-breaking. The mass of the states is determined
by the VEV $v$ and an arbitrary coupling constant $g$.
However, the parity violation appears naturally through the
gauging of the group with the help of the non-canonical
Pauli-$\sigma$-operators \cite{Deh95,Gei97}. Moreover, in the LQG,
such parity violation is described by matching the \emph{immirzi}
parameter (which measures the size of the quantum of area in
Planck units) with the black hole entropy \cite{Ran06}.\\
In general, the simplest way to generate the spontaneous breakdown
of symmetry is to have a Lagrangian with the Higgs potential
$V(\phi)$ and a transformed gauged field $\phi'_a=
U\phi_a=e^{\tilde{\lambda}^a \tau_a}=e^{\chi_a}$ in the following
form,
\begin{equation}
{\cal L}_H= {\cal L}(\phi)= \frac{1}{2}\, \phi^\dagger_{;\nu}\,
\phi^{;\nu} - V(\phi) = {\cal L}(\phi'),\label{L}
\end{equation}
where $;\mu \equiv D_\mu$ is the covariant derivative and
the potential $V(\phi)$ has the form as follows,
\begin{align}
V(\phi)= \frac{\mu^2}{2} \,\phi^\dagger \phi + \frac{\lambda}{4!}
\, (\phi^\dagger \phi)^2 \,\,,\label{Higgspot}
\end{align}
where $\mu^2<0$ and $\lambda>0$. Such theories are called
$\phi^4$-theories. The last term in the potential (\ref{Higgspot})
is not bilinear and it is crucial for the apparent symmetry
breakdown. The Lagrangian given by equation (\ref{L}) is invariant
under the spatial-inversion (i.e. $\phi \rightarrow -\phi$) with the
features of the tachyonic condensation (i.e. condensate for an
imaginary mass with $\mu^2 <0$). Such conditions are needed to stay
with the Higgs mode, which otherwise becomes a Wigner mode with
classical vacuum where self-interactions lack to produce the
necessary Higgs Mechanism at the relatively low energies of the
hodiernal universe. However, with the possibility of the tachyonic
condensation, the ground state $\phi_0$ becomes twice degenerate and
$\phi_z=0$ has a maximal value for the energy-density $u$. The
minimum energy is then given by the non-vanishing Higgs ground state
value (i.e. $v\neq 0$) in the following form:
\vspace{-0.2cm}
$$
u_0= u(\phi_0)= -\frac{3}{2} \, \frac{\mu^4}{\lambda} \equiv
u_{min},
$$
\vspace{-0.5cm}
\begin{equation}
\phi_0^{(+)}= \sqrt{-\frac{6\mu^2}{\lambda}} \, e^{i\alpha}
\equiv \tilde{v}= v\, e^{i \alpha} \neq 0.
\end{equation}
For a pure scalar case, $v$ is to be chosen between $\phi_0^{(-)}$
and $\phi_0^{(+)}$. As such, in technical jargons, the circle of
the localized minima for the minimality condition of $u$ is
popularly known as a \emph{Mexican hat} and the regions with
different $\phi_0$-values are called the \emph{topological
defects} while those changing with the values
$\phi=v\leftrightarrow -v$ are termed as \emph{interface domains}.
Further, it is also possible to have the choice of $\alpha=0$
without making any restriction to the system since this does not
demand any kind of physical changes. However, this choice does not
allow mass to go through the phase transitions without changing
its vacuum value. Therefore, even if the Lagrangian is invariant
under phase transitions, it must suffer the loss of invariance
explicitly through its ground state, and the particles that fall
in this state interact with the Higgs bosons and slow down. In
particular, in view of the Special Relativity (SR), the massless
particles travel with the speed of light $c$ and massive ones have
as speed $v<c$. So the mass generation of the particles may be
interpreted in relation to their interaction with the Higgs field.
In fact, the energy of the system is nothing but a meager and
$\phi$ lies near the minimum of energy. It is, therefore, possible
to expand the scalar field around its minimal state with its
excited values $\hat{\phi}$ in the following form:
\begin{equation}
\phi=v+\hat{\phi}.
\end{equation}
The Lagrangian (\ref{L}) may now be given in iso-scalar form (only
up to second order terms) as follows,
\begin{equation}
{\cal L}(\hat{\phi}) =
\frac{1}{2}\,\hat{\phi}^\dagger_{,\nu}\, \hat{\phi}^{,\nu} -
\frac{M_{H}^2}{2} \, \hat{\phi}^2 - \frac{\lambda}{3!} \, v
\, \hat{\phi}^3 - \frac{\lambda}{4!} \, \hat{\phi}^4  \, \neq  \,{\cal
L}(-\hat{\phi}).\label{Lexc}
\end{equation}
The first term in the Lagrangian (\ref{Lexc}) corresponds to
the kinetic energy of the Higgs field while the second one
represents the mass term (i.e. $ M_{H}^2 \equiv -2\mu^2$) for the
Higgs field. In fact, due to the presence of the term for the
excited field (i.e. $\hat{\phi}^3$) in the Lagrangian (\ref{Lexc}),
the symmetry is suddenly broken since the Lagrangian (\ref{Lexc}) is
not spatially invariant anymore. However, the Lagrangian in the
iso-vectorial form may be re-written as
\begin{widetext}
\begin{equation}
{\cal L}(\hat{\phi}) = \frac{1}{2} \, \hat{\phi}^{\dagger
a}\,_{,\nu} \, \, \hat{\phi}_a\,^{,\nu} + \frac{1}{2} \, g^2 A_{\mu a} \,^b\, \phi_0^{\dagger a}\, A^{\mu}\,_b \,^c \, \phi_{0 c} -
\frac{\lambda}{4!}\,(\phi_0 ^{\dagger a} \, \hat{\phi}_a  +
 \hat{\phi}^{\dagger a}  \, \phi_{0 a})^2 ,\label{Lexc-vec}
\end{equation}
\end{widetext}
where $\phi_a$ is a scalar quantity in spin-space. The Lagrangian in
iso-vectorial form is invariant under local transformations for
which it is needed to define the covariant derivative (different for
left- and right-handed states because of the couplings which give
the parity violation). As such, the massive term of the excited
Lagrangian with these excited Higgs scalar field leads to the term of the mass of the gauge bosons as
\begin{equation}
M_{A} = \frac{1}{2}\, g^2 \, A_{\mu a}\,^b \phi_0^{\dagger a}
A^\mu\,_b \,^c \phi_{0c} \sim ({\cal M}^2)^{ij}A_{\mu i}A^{\mu}\,_j,
\label{A-M2}
\end{equation}
where the mass term for the gauge boson $A_\mu$ comes from the
covariant derivative, and the mass-square matrix (operator), which
is symmetric and real, is given below in the natural system of
units:
\begin{equation}
({\cal M}^2)^{ij}= 4\pi g^2 \phi_0^\dagger \tau^{(i}\tau^{j)}
\phi_0  = 2\pi g^2 v^2 (c^{ij} \underline{1} + N^\dagger d^{ij}\,_k \tau^k N).\label{M2}
\end{equation}
Such broken phase of symmetry cannot be reached by perturbative
expansion techniques from the normal vacuum. The SSB may therefore
be thought of as a phase transition which is manifestly
non-perturbative. However, the Higgs field does not give mass to
the neutrinos in the usual form of SM. In the equation (\ref{M2}),
$\tau^i=\tau^{i\, \dagger}$ are the generators of the gauge group
and they satisfy the following properties for the rest and broken
phase of the symmetry respectively:
\begin{equation} \left.
\begin{array}{c}
\tau^j_a\, ^b N_b=0 \\
\tau^j_a\, ^b N_b \neq 0
\end{array}
\right\} \,.
\end{equation}
The diagonal components of the mass-square matrix are positive
definite and correspond to the mass of the gauge bosons coupled to
the scalar vector field $\phi$. The masses corresponding to the
Higgs scalar multiplet are then $M_{H}^2 = - 2\mu^2$ and  $M_{G} =
0$ where the massless Goldstone boson $(M_G)$ belongs to the
Nambu-Goldstone mode because of the global symmetry breakdown which
carries the quantum number of the broken generator. Further, with
the conditions of conserved current corresponding to an exact
symmetry of the Lagrangian, the non-invariant vacuum follows
$\chi_a|0> \neq 0$ (where $\chi_a$ is the Goldstone boson field)
while Lorenz invariance implies that
$\tilde{\lambda}^a\tau_a=\chi_a$ is valid at least for one $a$.
Moreover, there must be a state $|m>\in {\cal H}$ with $<m|\chi_a|0>
\neq 0$ for a massless spin-$0$ particle. Nevertheless, in nature
such massless scalar spin-$0$ particles do not seem to exist and the
Goldstone bosons would give rise to long-ranged forces in classical
physics to generate new effects in various scattering and decay
processes in nature. The possible non-relativistic long-range forces
arising from the existence of massless Goldstone particles are
spin-dependent, and it is quite difficult to observe them directly.
However, in principle, the $\gamma_5$-couplings along with the CP-
violation would change to scalar interactions which may then
subsequently lead to spin-independent long-range forces
\cite{Moh86}. The existence of such Goldstone bosons may also affect
the astrophysical considerations with some sort of new mechanism for
the energy loss in stars. Furthermore, the excited Higgs field
distinguishes from the ground state by a local transformation that
can be gauged away through an inverse unitary transformation
$U^{-1}$. Such unitary transformations contain the Goldstone fields
$\tilde{\lambda}$ (as the generator of symmetry) in the following
form,
\begin{equation}
U=e^{\tilde{i\lambda}^a\tau_a}=e^{i\chi_a},
\end{equation}
and it is possible to gauge out the Goldstone bosons from the
theory by the following unitary gauge transformations,
\begin{equation}
\phi=\frac{\rho}{v} \, U \phi_0=\rho UN,
\end{equation}
and consequently we have,
\begin{equation}
\phi \rightarrow U^{-1}\phi =\varrho \, (U^{-1}U) N=\varrho\,
N;~~~~ \psi \rightarrow U^{-1}\psi,
\end{equation}
with $\rho=\phi^\dagger \phi$, $\phi=v(1+\varphi)=v \,\zeta$. The
absence of the Goldstone bosons is then mathematically permitted,
which indicates that Goldstone's rule of massless particles in the
broken phase of symmetry is only valid for the global gauge while
the unitary gauge considered here is a local one. In SM, the
gauge fixing for the leptonic multiplet is given by $N=(0,1)^T$ with
$SU(2)_ W\otimes U(1)_Y$ as followed from the electro-weak
interactions,
\begin{equation}
\left.
\begin{array}{c}
\psi_{fL}= {{\nu_f} \choose {e_f}}_L\\
\psi_{fR}=e_{fR}
\end{array}
\right\} \,,\label{psi}
\end{equation}
where the parity is defined by the projection operator $(1\pm
\gamma^5)$ . The $+$ and $-$ signatures denote the left $(L)$ and
right $(R)$-handedness of the particles respectively. In equation
(\ref{psi}), $f$ represents the family of leptons i.e. $f=$($e$,
$\mu$, $\tau$). For instance, the masses of the first generation
leptons (i.e. electron and its corresponding neutrino) are given
as follows,
\begin{equation}
\left.
\begin{array}{c}
 M_e= G \, v\\ M_\nu =0\label{me}
\end{array}
\right\} \,,
\end{equation}
while the masses of the gauge-bosons are defined as below,
\begin{equation}
\left.
\begin{array}{l}
M_W=\sqrt{\pi} g_2 v= \left[\frac{\pi \alpha}{\sqrt 2\,G_F \sin^2
\vartheta_w}\right]^{\frac{1}{2}} = \frac{37.3
GeV}{\sin\vartheta_w}\\ \\ \quad M_Z=\frac{M_W}{c_w} > \, M_W
\end{array}
\right\} \,,
\end{equation}
where $G_F \simeq 1.166 \times 10^{-5}$GeV$^{-2}$ and $\alpha
\simeq 1/137$ are the Fermi and Sommerfeld structure constants,
respectively. However, the Weinberg term $c_w = \cos \vartheta_w$
(for the mixing of $Z^0$ with the $W^\pm$ and $A$) may be defined in
terms of the coupling constant of the hypercharge in the following
form,
\begin{equation}
g_2 \sin \vartheta_w = g_1 \cos \vartheta_w = e.
\end{equation}
The measurements for the Weinberg mixing angle $(\vartheta_w)$
within SM lead to the following approximate values,
\begin{equation}
\left.
\begin{array}{c}
\sin^2 \vartheta_w \cong 0\cdot23 \\
\cos^2 \vartheta_w \cong 0\cdot77
\end{array}
\right\} \,.
\end{equation}
The Weinberg mixing angle not only relates the masses of $W^{\pm}$
and $Z^0$ bosons, but also gives a relationship among the
electromagnetic $(e)$, charged weak $(g_2)$ and neutral $(g_1)$
couplings and ultimately leads to the following approximate values
of mass for $W$ and $Z$ bosons,
\begin{equation}
\left.
\begin{array}{c}
M_W \cong 78 \, \text{GeV} \\
M_Z \cong 89 \, \text{GeV}
\end{array}
\right\} \,.
\end{equation}
However, the baryonic matter field is given in terms of the doublets
where $f$ denotes the different generations of quarks (while in QCD,
it counts the flavor with the color-triplet of $SU(3)_C$). This
interaction is given by a new enlargement term in the
Lagrangian (which is necessary to generate the mass of fermions via
Higgs Mechanism) as below,
$$
{\cal L}(\phi, \psi)=  - G_f \,(\bar{\psi}^A \phi^{\dagger
a}\hat{x}\psi_{aA}+ \bar{\psi}^{aA} \hat{x}^\dagger \phi_a \psi_A)
$$
\begin{equation}
 \equiv  - M_f(\bar{\psi}^AN^{\dagger a} \psi_{aA}+ \bar{\psi}^{aA}N_a
 \psi_A).
\end{equation}
The propagator for the exchanged boson (i.e. Higgs boson) via the
Higgs interaction of two fermions turns out to be in the lowest order
of the amplitude the same as derived from a Yukawa
potential (i.e. a screened Coulomb potential). The propagator or
Green function of such Klein-Gordon equation of a massive particle
itself is enough to demonstrate that the Higgs interaction is of
Yukawa-type. In fact, the scalar field $(\phi_a)$ couples with
fermions $(\psi_A)$ through the Yukawa matrix $\hat{x}$, and the
mass of the fermions may be then given as $M_f=G_f \, v$, and it
is worth to notice that the equation (\ref{me}) is a special case
of it. Such Higgs-coupling to the fermions is model-dependent,
although their form is often constrained by discrete symmetries
imposed in order to avoid 3-level flavor changing neutral currents
mediated by the Higgs exchange. However, to have a more accurate
picture, the quantum mechanical radiative corrections are to be
added in order to have an effective potential $V_{eff}(\phi)$.
Since the coupling is also dependent on the effective mass of the
field, the $\lambda \, \mu^2 \phi^2$ and $\lambda^2 \phi^4$ terms
from a vacuum energy contribution are caused by vacuum
fluctuations of the $\phi$-field and must be incorporated in the
system to have a correct physical description. Furthermore, there
are additional quantum gravitational contributions and temperature
dependence so that $V_{eff}(\phi) \rightarrow V_{eff}(\phi, T)
\sim V_{eff}(\phi)+ M^2(\phi) \,T^2- T^4$. As a consequence,
symmetry must be restored at high energies (or temperatures),
especially in the primordial universe \cite{Cer96}, which is
contrary to the present state of the universe. The symmetry
breakdown through the cooling of the universe after the Big Bang,
in turn, provokes the appearance of the four well-known elementary
interactions. In this context, though, it is an open question if
there are more than one Higgs particle; it would be necessary that
at least two of them exist in usual unifying theories to occur SSB
so that gauge bosons become massive at different energy-scales as
in GUT. Moreover, there is the symmetry breakdown of
parity, too (which may be understood in terms of \emph{axions})
\cite{Pec77, Wei78}, the mechanism of which was claimed experimentally demonstrated shortly sometimes earlier during the year 2006 \cite{Zav06}.
\section{Some phenomenological aspects}
\noindent Though the SM explains and foresaw many aspects of nature
proven by the well tested experiments, there is still a problem of
special relevance which is popularly known as the {\it hierarchy} problem. The EW breaking scale related to the Higgs mass is expected too high in SM, and this seems quite unnatural to many physicists. This problem has apparently no solution within SM. If it can be solved, it can then signify non-elementarity of the Higgs fields. Indeed, if they
are elementary, then there must be a symmetry protecting such fields
from a large radiative correction to their masses \cite{Kan97}. In
order to do so, the first choice is to take the Higgs field as a
composite structure containing only an effective field (in the way
one can explain superconductivity as following Higgs Mechanism), and
it then seems indeed possible to construct a renormalizable SM
without the fundamental Higgs scalar field \cite{Gie03}. However,
the second choice is to supersymmetrize the SM, which leads to the
possibility to cancel the unphysical quadratic divergences in the
theory; in this way, it is possible to provide an answer to the
{\it hierarchy} problem between the EW and Planck mass scale. The supersymmetric version of SM may, therefore, be an important tool to stabilize the {\it hierarchy} against the quantum corrections. As such, in the MSSM with the radiative EW symmetry breaking, the stability of the Higgs potential leads to the mass generation around the EW scale. In such SuSy versions one uses two doublets for the Higgs field as follows, \begin{equation}
\Phi_1={{\phi ^{0*}_1}
\choose {\phi ^{-}_1}} ; \,\, \, \Phi_2={{\phi^{+}_2} \choose
{\phi ^{0}_0}}\,,
\end{equation}
to generate the mass of up and down fermions. The most general potential \cite{Kan93} for this purpose is of the following form,
\begin{widetext}
$$
V ( \Phi_1\, \Phi_2) = M_{11}^2 \Phi^\dagger_1 \Phi_1+ M_{22}^2
\Phi^\dagger_2 \Phi_2- [M_{12}^2 \Phi^\dagger_1 \Phi_2] +
\frac{1}{2}\lambda_1(\Phi_1^\dagger \Phi_1)^2 +
\frac{1}{2}\lambda_2(\Phi_2^\dagger \Phi_2)^2
$$
\begin{equation}
+ \lambda_3 (\Phi_1^\dagger \Phi_1)(\Phi_2^\dagger \Phi_2) + \lambda_4 (\Phi_1^\dagger \Phi_2)(\Phi_2^\dagger \Phi_1) +
\frac{1}{2}\lambda_5(\Phi_1^\dagger \Phi_2)^2 +
[\lambda_6(\Phi_1^\dagger \Phi_1)+ \lambda_7(\Phi_2^\dagger
\Phi_2)]+ h.c.\, ,
\end{equation}
\end{widetext}
where the $\lambda_6$ and $\lambda_7$ are often
dropped out in view of the cancelation by the following symmetry:
\begin{equation}
\Phi_1 \rightarrow -\Phi_1 \, \Rightarrow \, M_{12}=0.
\end{equation}
In view of the above-mentioned potential, the scalar field develops
a non-degenerate VEV (with $M_{ij}^2$ having at least one negative
eigenvalue). The minimization of the potential leads to
\begin{equation}
<\Phi_1>= {{0} \choose {v_1}}; \,\,\, <\Phi_2>= {{0} \choose
{v_2}},
\end{equation}
which in turn defines,
\begin{equation}
v^2= v_1^2+ v_2^2=\frac{4M_W^2}{g^2}=(246\,\text{GeV})^2;~~~~
\tan\beta= \frac{v_2}{v_1}\,.
\end{equation}
In this scenario, there are eight degrees of freedom in total
including three Goldstone bosons ($G^{\pm}, G^0$) those are absorbed
by $W^\pm$ and $Z^0$ bosons. The remaining physical Higgs particles
are two CP-even scalar particles ($h^0$ and $H^0$ with $M_{h^0}
<M_{H^0}$) (one is a CP-odd scalar $A^0$ and the other is a charged
Higgs pair $H^\pm$). In fact, two of the neutral fields come from
the real part $\Re e\phi_1^0$ and $\Re e\phi_2^0$ and the third
actually belongs to the imaginary part of a linear combination of
$\Phi_1$ and $\Phi_2$ \cite{Moh86}. However, the mass parameters
$M_{11}$ and $M_{22}$ can be eliminated by minimizing the scalar
potential. The resulting squared masses for the CP-odd and charged
Higgs states are then given as follows:
\begin{alignat}{1}
M_{A^0}^2=& \frac{2\, M_{12}^2}{\sin 2 \beta} - \frac{1}{2} v^2(2
\lambda_5+ \lambda_6 \tan^{-1}\beta+ \lambda_7 \tan\beta),\\
M_{H^\pm}^2=& M_{A^0}^2+ \frac{1}{2}v^2(\lambda_5- \lambda_4).
\end{alignat}
The SuSy, therefore, couples the fermions and bosons in such a way that
the scalar masses have two sources (given by $v$) for their
quadratic divergences, one from scalar loop which comes with a
positive sign and another from a fermion loop with negative
sign. The radiative corrections to the scalar masses can be
controlled by canceling the contributions of the particles and
their SuSy partners (i.e. s-particles) those come in the spectrum
because of the breakdown of SuSy. Since SuSy is not exact, such
cancelation might not be complete as the Higgs mass receives a
contribution from the correction which is limited by the extent of
SuSy-breaking. However, in the structure of this model, the
quantum loop corrections would induce the symmetry-breaking in a
natural way, and they may be helpful in solving some other
conceptual problems of SM \cite{Kan93}. Since the appearance of
the top-quark at an extremely high energy-scale (in comparison to
the one of all other quark flavors) could not be explained within
the well established notions of SM, it might be understood as
consequence of an unknown substructure of the theory. Therefore,
the SM and MSSM would be only an effective field theory with
another gauge force which is strong at $SU(2) \otimes U(1)$
breaking-scale. On the other hand, in technicolor (TC) theory, the
Higgs particles are not believed to be fundamental and the
introduction of technifermions (a type of pre-quarks or preons)
represent the quarks as composite particles with a new symmetry
which is spontaneously broken when technifermions develop a
dynamical mass (independently of any external or fundamental
scalar fields. The developments of the TC theory, however, has
not yet been fully able to suppress the possibility of such scalar
fields completely out, so that an extended version of it would be
required (i.e. ETC) \cite{Kan93}. Furthermore, another interesting fact
is the heavy mass of the top-quarks, which leads to some models
claiming that such a mass may only be generated dynamically with
so-called top condensation \cite{Kan97} where the Higgs particles
might have a phenomenological presence with the $t\bar{t}$ bound
states of top-quarks \cite{Kan93, Bij94}). Next, and listing more,
if one thinks about the existence of Higgs particles in terms of
the developments in LQG, the existence of the Higgs particles can
be questioned because in LQG the particles are derived though a
preon-inspired (Helon) model as excitations of the discrete
space-time. Further, the Helon model does not offer a \emph{preon}
for Higgs (as a \emph{braid} of space-time) so that it lacks of
the same until the model indeed finds an expansion with a Higgs \emph{braid} \cite{Bil06}. On the other hand, the phenomenological
nature of Higgs (which might also explain other problems as the
impossibility to measure the gravitational constant $G$ exactly
\cite{Fae-Deh05}) might also be a consequence of a more profound
coupling as with gravitation. In view of the fact that the Higgs
particles couple gravitationally within the SM
\cite{Bij95,Deh91,Deh90}, the consequences are discussed in
detail in the next section.
\section{The Gravitational-like interactions and Higgs mechanism}
\noindent The general relativistic models with a scalar field
coupled to the tensor field of GR are conformally equivalent to
the multi-dimensional models, and using Jordan's isomorphy
theorem, the projective spaces (like in the Kaluza-Klein's theory)
may be reduced to the usual Riemannian $4-dim$ spaces
\cite{Fau01}. Such scalar field in the metric first appeared
in Jordan's theory \cite{Jor55} and manifests itself as
\emph{dilaton}, \emph{radion} or \emph{gravi-scalar} for different
cases, and those in fact correspond to a scalar field added to GR
in a particular model \cite{Cot97}. The gravitational constant $G$ is then replaced by the reciprocal of the average value of a scalar field through which the strength of gravity can be varied (thus breaking the
SEP), as was first introduced by Brans and Dicke \cite{Bra61} by
coupling a scalar field with the curvature scalar in the Lagrangian $\cal{L}$.\\
However, a more general covariant theory of gravitation can
accommodate a massive scalar field in addition to the massless
tensor field \cite{OHan72, Ach73} so that a generalized version of
the Jordan-Brans-Dicke (JBD) theory with massive scalar fields can
be derived \cite{Fuj74}. It is worth mentioning that Zee was the
first who incorporated the concept of SSB in the
STT of gravitation \cite{Zee79}. It
represents a special case of the so-called \emph{Bergmann-Wagoner}
class of STTs \cite{Berg68, Bro01}. The latter is more general
than that of the JBD class alone (where $\omega=const$ and
$\Lambda(\phi)=0$), because of the dependence of the coupling
term $\omega$ on the scalar field and of an appearing cosmological function. In Zee's approach, the function $\Lambda(\phi)$ depends on a symmetry breaking potential $V(\phi)$, and it is therefore quite reasonable to consider a coupling of Higgs particles with those which acquire mass through a short-ranged gravitational-like interaction within SM \cite{Bij95, Deh91,Deh90}. Such a model is compatible
with Einstein's ideas to the Mach principle \cite{Ein13}.\\
The simplest Higgs field model beyond the SM consists of a single
particle which only interacts with the Higgs sector of SM.
With a fundamental gauge-invariant construction block
$\phi^\dagger \phi$, the simplest coupling of a particle to the
Higgs field may be defined as $\breve{\lambda} \phi^\dagger \phi X
$ where $X$ is a scalar field. The Higgs field develops a VEV and,
after shifting it, the vertex leads to a mixing between the scalar
and the Higgs field, which may give rise to new effects those do not
involve the scalar explicitly \cite{Hil87}. The X-field may not be
considered as fundamental, but an effective description of an
underlying dynamical mechanism is possible through its connection
to the technicolor theories \cite{Hil87} (i.e. alternatively to a
connection between the gravity and Higgs sector). In fact, both
the gravity and Higgs particles possess some universal
characteristics, and such a commonality leads to a relation between
the Higgs sector and gravity which is popularly termed as the
Higgs field gravity \cite{Deh91}. Further, there may be a
similarity between $X$ and the hypothetical graviton since both
are the singlets under the gauge group \cite{Bij95}. They have
no coupling to the ordinary matter and therefore have
experimental constraints for their observations. One can even argue about their absence from the theory because they can have a bare mass term which can be made to be of the order of the Planck mass and that makes these fields invisible. However, one can assume that all the masses including that of the Planck mass are given by SSB processes in
nature. In this case, there is a {\it hierarchy} of mass scales $M_P\gg
v$. With these similarities, $X$ can be considered to be
essentially the graviton and may be identified with the curvature
scalar ${\cal R}$ \cite{Bij95}. Moreover, this possibility may be used to explain the naturalness problem, especially since other candidates
as top quark condensation or technicolor have not functioned well
so far and supersymmetry doubles the spectrum of elementary
particles replacing Bose (Fermi) degrees of freedom with Fermi
(Bose) degrees of freedom and with all supersymmetric particles which are by now beyond physical reality. However, the cut-off of the theory at which the Higgs mass is expected may not be so large and of the
order of the weak scale \cite{Bij94}. The Higgs particles
therefore seem to couple naturally to the gravitation, and a STT of
gravitation with a general form of Higgs field for symmetry
breaking can indeed be derived within SM  \cite{Deh92, Deh93}.
Moreover, Higgs may be explained as a phenomenological appearance
of the polarization of the vacuum since it leads to a cosmological
term  which may be identified in terms of the Higgs potential with a functional coupling parameter $G$ \cite{Deh92,Deh93}. In such STTs the scalar field $\phi$ may behave similarly to a cosmon \cite{Bij94}. The Higgs field may, therefore, also contribute in cosmological range as a part of the Cold Dark Matter (CDM) because of the functional nature of the coupling $G$ and the self-interacting DM (SIDM)
\cite{Ges92}-\cite{Bez07b}. However, the unification of
gravitation with the SM and GUT using Higgs field may also
explain Inflation for baryogenesis and solve the flatness problem.
The scalar fields and Higgs Mechanism lead to various inflationary
models in cosmology  where the
cosmological constant produces the inflationary expansion of the
universe \cite{Per98}-\cite{Sch79}. Within the original,
\emph{old} inflation \cite{Gut81},
the scalar field should tunnel from its false vacuum to the
minimal value $v$ while, on the other hand, in the \emph{new}
inflation it rolls slowly from $\phi \ll v$ to $v$ and then
oscillates near to it. However, in case of the \emph{chaotic}
inflation, the rolling-over is explained in the range from $\phi
\gg v$ to $v$ \cite{Cer96}. The new inflation can lead in a
symmetry-broken STT to a deflation epoch before the expansion, and
then a fine-tuning is needed for the universe not to collapse in a
singularity again. However, for the chaotic inflation, which seems
to be the most natural form of inflation not only within theories
of induced gravity with Higgs fields but in general models
\cite{Lin05}, too, a singularity at the beginning of time is not
needed because of a breaking of the Hawking-Penrose energy
condition \cite{Pen65}-\cite{Haw68}. In fact, this is possible for sufficiently large negative pressures which are possible as a consequence of Yukawa-type interactions \cite{Lib69} that might play an important role in early stages of the universe \cite{Deh75}. Therefore, the chaotic inflation is preluded in general not by a singularity of a
Big Bang but by a so-called Big Bounce, having its signatures
within LQG \cite{Ash06}. In particular, in a theory of induced
gravity with Higgs mechanism \cite{Cer95b}, after
inflation, the Higgs potential might decay in to the baryons and leptons with the oscillations from the Higgs potential which might be
interpreted in terms of the SIDM in form of the Higgs particles.
The actual interactions of Higgs particles are only given by
their coupling to the particles within SM and the field equation for
the Higgs field \cite{Deh92} can be given as follows,
\begin{align}
\phi_{;\mu}\,^{;\mu} +\mu^2\phi+ \frac{\lambda}{6}(\phi^\dagger
\phi) \, \phi= -2G \, (\,\bar{\psi}_{R}\hat{x} \psi_L) ,
\end{align}
where $\hat{x}$ is the Yukawa coupling operator which represents the coupling of the Higgs field to the fermions, and the subscripts $L$ and $R$ refer the left- and right-handed fermionic states of $\psi$ respectively. In SM the source of the Higgs field are the particles that acquire mass through it, and the Lagrangian for the case of a coupling of the Higgs field to space-time curvature through the Ricci scalar
$({\cal R})$  \cite{Deh92} is given in the following form,
\begin{alignat}{1}
{\cal L}= [\, \frac{\breve{\alpha}}{16\pi}\, \phi^\dagger \phi\,  {\cal R} +
\frac{1}{2}\, \phi^\dagger_{;\mu} \, \phi^{;\mu}- V(\phi)\, ] + {\cal L}_M
\, , \label{LHSTT}
\end{alignat}
where the dimensionless gravitational coupling parameter $\breve{\alpha}$ can be interpreted as a remnant of a very strong interaction which is given
by the ratio of Planck's mass to boson mass as $\breve{\alpha}
\simeq (M_P/M_B)^2\gg 1$ \cite{Bij95}. On the other hand,
$\breve{\alpha}$ is coupled to the gravitational strength $G$, on
which the redefined scalar field mass of the model (i.e. $M_H$) is
dependent. This mass is expected around a $10^{-17}$ part of the
value needed in SM  and $10^{-4}$ of the one in GUT under SU(5). It
is even possible for the Higgs particles to decouple from the rest
of the universe and interact only gravitationally. That is the case
if the same $\phi$ has a coupling to ${\cal R}$ in ${\cal L}_M$ for mass generation of the gauge bosons \cite{Deh93}. Within this scalar-tensor theory then, the Higgs field equations with coupling to ${\cal R}$
and $\phi$ ($\sim$ SM) and only to ${\cal R} $ ($\sim$ GUT), respectively, are given below:
\begin{alignat}{1}
\phi^{;\mu}\,_{;\mu}- \frac{ \breve{\alpha}}{8\pi} \, {\cal R} \, \phi +
\mu^2\phi+ \frac{\lambda}{6}\, (\phi^\dagger \phi) \, \phi&= -2G\,
(\bar{\psi}_R \hat{x}
\psi_L) \,,\label{kphi}\\
 \phi^{;\mu}\,_{;\mu}- \frac{ \breve{\alpha}}{8\pi}  \, {\cal R} \, \phi + \mu^2\phi+
\frac{\lambda}{6} \, (\phi^\dagger \phi) \,  \phi & = 0.
\end{alignat}
The field equation (\ref{kphi}) after the symmetry breakdown (with the excited Higgs field which satisfies $(1+\varphi)^2= 1+\xi $) acquires the following form,
\begin{equation}
\xi^{,\mu}\,_{;\mu}+ M^{*2}\xi=\frac{1}{1+4\pi/3\breve{\alpha}}\,
\frac{8\pi \tilde{G}}{3} \, [\,T - \sqrt{1+\xi} \,\,  \bar{\psi}\hat{m}\psi
\,],\label{trace}
\end{equation}
where $\tilde{G}=({\breve{\alpha} v^2})^{-1}$ is related to Newton's constant. The Higgs field mass is given by
\begin{equation}
M^{*2}= l^{-2}= \left[ { \frac{16 \pi \tilde{G} ( \mu^4 / \lambda ) }{1 + \frac {4 \pi }{3 \breve{\alpha} } }  }
\right] = \left[\frac{\frac{4\pi}{9\breve{\alpha}}\lambda v^2}{1+
\frac{4\pi}{9\breve{\alpha}} }\right] , \label{M}
\end{equation}
which indicates that the Higgs field possesses a finite range
defined by the length scale $l$. $T$ is the trace of the
energy-stress-tensor $T_{\mu \nu}$ given in the following form,
\begin{equation}
T^{\mu \nu} = \frac{i}{2} \, \bar{\psi}\gamma^{(\mu}_{L,R}\psi^{;\nu)}+ h.c.-
\frac{1}{4\pi} \, (F^{\mu}\,_\lambda^a F^{\nu \lambda}_a -
F^a_{\alpha \beta}F_a^{\alpha \beta}g^{\mu \nu})\,. \label{FT}
\end{equation}
The field-strength tensor in equation (\ref{FT}) is defined as
$F_{\mu \nu}=-ig^{-1}\, [{\cal D}_\mu,{\cal D}_\nu]$ where ${\cal D_\mu}$
is the covariant derivative. However, the generalized Dirac matrices $\gamma^\mu=h_a^\mu \gamma^a$ in equation (\ref{FT}) satisfy the following relation,
\begin{align}
\gamma^\mu \gamma^\nu + \gamma^\nu \gamma^\mu=2 \, g^{\mu \nu} \, \underline{1}\,.
\end{align}
The trace of the energy-stress tensor as mentioned in equation (\ref{trace}) is then given as follows,
\begin{align}
T=\frac{i}{2}\bar{\psi}\gamma^\mu_{L,R}\psi_{;\mu}+ h.c.=
\sqrt{1+\xi} \,  \,\bar{\psi}\hat{m}\psi.
\end{align}
However, in view of the coupling of $\phi$ with the
matter-Lagrangian, the energy-stress tensor and
source of Higgs particles cancel the contribution due to each other
such that the Higgs particles are no longer able to be generated and
interact only through the gravitational channel.
\section{Search for Higgs boson and constrains}
\noindent The search for the Higgs boson is the premier goal for the
high energy physicists as the SM without the Higgs boson (or at
least Higgs Mechanism) is not manifestly consistent with nature. It
is often said that the Higgs boson is the only missing piece of SM
as the top-quarks are now already subject of experimental reality. The
Higgs bosons could not be generated so far in the particle
accelerators, although their practical reality in explaining mass is
not questioned at all and has been proved in many ways. However, the
fundamental character of Higgs bosons still demands more explanation,
and due to the absence of experimental evidence, they lie in the
category of yet to be discovered objects. From the point of view of
the search for Higgs particles, SuSy-models are leading candidates
(although no supersymmetric particles have been discovered so far),
while TC-models do not contain Higgs particles at all and some
gravitational theories are often interpreted as with Higgs particles
only interacting gravitationally. The MSSM, having the particle
spectrum of SM along with the corresponding superpartners and two
Higgs doublets in order to produce mass, is consistent with
supersymmetry as well avoiding the gauge anomalies due to the
fermionic superpartners of bosons, stabilizing the search for Higgs
mass. Using group renormalization techniques, MSSM Higgs masses have
been calculated, and two specific bounds on it can be made for the
case when the top-squark mixing is almost negligible and for the
case when it is maximal. The assumption $M_T = 175$GeV and
$M_{\overline T} = 1$ TeV leads to $M_{h^0} \gtrsim 112$ GeV when the
mixing is negligible, while maximal mixing produces the large value
of $M_{h^0} \gtrsim 125$ GeV. On the other hand, within a MSSM with
explicit CP violation, to constrain CP phases in the MSSM, the
measurements of thallium electric dipole moments (DPM) are used
\cite{Lee07}. The present experimental constraints suggest that the
lightest Higgs mass (say $H_1$) has to lie in the range 7 GeV $\lesssim
M_{H1}\lesssim 7.5$ GeV ($\tan \beta \backsimeq 3$), or $\lesssim
10$GeV ($3\lesssim \tan \beta \lesssim 5$), assuming mild
cancelations in the thallium EDM. In a scenario with explicit CP
violation in MSSM, the lightest Higgs boson can be very light
($M_{H1}\lesssim 10$GeV), with the other two neutral Higgs bosons
significantly heavier ($M_{H2,H3}\gtrsim 100$GeV). Here, CP is
explicitly broken at the loop level, and the three neutral MSSM Higgs
mass eigenstates have no longer CP parities due to a CP-violating
mixing between the scalar and pseudo-scalar neutral Higgs bosons.
The lightest Higgs boson is mostly CP odd, and its production
possibility at the Large Electron-Positron Collider (LEP) is highly
suppressed. The second-lightest Higgs ($H_2$) at $\sim 110$ GeV
dominantly decays in $H_1$, which then decays in two $b$ quarks and
$\tau$ leptons. This leads to a decay mode containing 6 jets in the
final state \cite{Lee07} that was recovered with only low efficiency
by the LEP2.\\
In the search for the Higgs boson, the mass of Higgs particles is a
constraint, and searches for it started in the early 1980's with the
LEP1, and without knowing all parameters, it was nearly impossible
to know at which energy-scale to search. In this constraint, maximal
mixing corresponds to an off-diagonal squark squared-mass that
produces the largest value $M_{h^0}$ with extremely large splitting
in top-squark mass eigenstates. On the other hand, the weak scale
SuSy predicts $M_h \gtrsim 130$ GeV, all relatively in accordance
with a possible Higgs mass of the order $114$GeV for which CERN
presented possible positive results in September 2000 (this was
achieved after delaying the shut-down of LEP and reducing the number
of collisions for the Higgs search to get additional energy and work
over the original capacities of the collider). However, the
experiments were forced to stop for further improvement in the
accelerator (towards the new Lepton-Hadron Collider (LHC)) and such
results could not be achieved once again by other laboratory groups.
Further, the updates presented in 2001 lessened the confidence and
more thorough analysis reduced the statistical significance of the
data to almost nothing. However, with $M_T$ known, at least another
parameter is given in the theory, i.e. $\tan \beta
={v_2} / {v_1}\thickapprox 1$ for all SuSy energy-scales
\cite{Kan93}. Here, $v_i$, $i=1,2$ are the ground state values of
each Higgs doublet needed for SuSy. In fact, the search for Higgs
uses different possible decaying processes, and especially in the
LEP, the decaying processes of electron-positron collisions produce
$WW$, $ZZ$ and $\gamma \gamma$-pairs in most of the cases, given in
the following form \cite{Wel03},
\begin{equation}
\left.
\begin{array}{c}
e^+e^-\rightarrow W^+W^-\\
e^+e^-\rightarrow ZZ\\
e^+e^-\rightarrow W^+W^-\gamma\\
e^+e^-\rightarrow \gamma \gamma
\end{array}
\right\} \,.
\end{equation}
There are also other possible channels where hadrons or heavier
lepton-pairs are seen in $e^+e^-$ collisions as given below,
\begin{equation}
\left.
\begin{array}{c}
e^+e^-\rightarrow e^+e^- q\bar{q}\\
e^+e^-\rightarrow q\bar{q}(\gamma)\\
e^+e^-\rightarrow \mu^+\mu^-
\end{array}
\right\} \,.
\end{equation}
Nevertheless, in experiments searching for Higgs particles, it is
important to separate them from the $HZ$-channel (i.e. $e^+e^-
\rightarrow HZ$), and for this one has to pick out the $H$ and $Z$
decay products against the background of all other decay channels,
although the cross-section is very small for the $HZ$ channel with
respect to hadron ones, and smaller for the greater masses of the
Higgs particles. Especially the $ZZ$ production is an irreducible
background to $ZH$ production. The $Z$ bosons can decay in $W$
bosons or, from an excited state, in other $Z$ and Higgs bosons $H$.
Then again, $H$ is expected to decay into four jets with 60\%
possibility in the form of heavy hadrons,
\begin{equation}
\left.
\begin{array}{c}
H \rightarrow b\bar{b} \\
Z\rightarrow q\bar{q}
\end{array}
\right\} \,.
\end{equation}
There is missing energy with 18\% possibility, while a leptonic
channel can exists having 6\% possibility,
\begin{equation}
\left.
\begin{array}{c}
H \rightarrow b\bar{b} \\
Z\rightarrow l^+l^-
\end{array}
\right\} \,.
\end{equation}
Moreover, another channel, which has 9\% possibility, is the
$\tau$-channel,
\begin{equation}
\left.
\begin{array}{c}
H \rightarrow b\bar{b}(\tau^+ \tau^-) \\
Z\rightarrow \tau^+ \tau^-(q\bar{q})
\end{array}
\right\} \,.
\end{equation}
Thus, experiments searching for $Z$ boson events are therefore
accompanied by a pair of bottom-quarks which all together have enough energy to come from a very heavy object (the Higgs boson candidates). Then, the total number of events in all decaying channels are to be compared against the total number expected in the theory along with the measured particle energy against the machine performance at all time-events to be sure that the changes in the accelerator energy and collision rate do not affect the interpretation. The cross-section for $HZ$ channels is meager and dependent on the mass of the Higgs particles. For a Higgs mass of $M_H=110$ GeV, the cross-section for
$e^+e^-\rightarrow HZ$-decays is already smaller than for $ZZ$ ones,
and decreasing for higher masses. For energies greater than
$110$ GeV, the cross-section of $e^+ e^- \rightarrow e^+ e^- q
\bar{q}$-decays is the biggest, in the scale of $10^4 pb$ and
increasing, while $e^+ e^- \rightarrow q \bar{q} (\gamma)$ is the
next one, around $10^3 -10^2 pb$ and increasing \cite{Wel03}. The
cross-sections of a decay in muons $\mu^+ \mu^- (\gamma)$ and in
$\gamma$-rays are almost equal. The decay channel in weakons $e^+
e^- \rightarrow W^+ W^-$ ascends rapidly in energies higher than
around $140$GeV, and it is almost constant after around $170$GeV with
a cross-section something larger than for $WW$ and $\gamma \gamma$ in a
cross-section scale of $10pb$. The decay channel in $ZZ$ is much
weaker but perceivably higher than zero, around $1pb$ after energies
in the scale of $180$GeV. The same for the shorter $W^+W^-\gamma$
decay channel. The $HZ$ channel is expected even weaker than the
last ones and perceivably unlike zero only after around $220$GeV, in
a cross-section scale of $10^{-1}pb$. For the decay channels to be
analyzed, it is important to detect the $Z$ decays. Further, OPAL
also detects $Z\rightarrow e^+e^-$ events. The electron pair events
have low multiplicity and electrons are identified by a track in the
central detector and a large energy deposit in the electromagnetic
calorimeter, $E/p = 1$. The $Z\rightarrow \mu^+\mu^-$ events are
also analyzed in L3 where the muons penetrate the entire detector
and let a small amount of energy in the calorimeters. The L3
emphasizes lepton and photon ID with a precise BGO crystal ECAL and
a large muon spectrometer. All detectors reside inside a r=6m
solenoid with a magnetic field B=0.5T. The $Z\rightarrow
\tau^+\tau^-$ events are also detected by the DELPHI collaboration.
Tau lepton-decays are dominated by 1 and 3 charged tracks, with or
without neutrals, missing neutrino(s) and back-to-back very narrow
\emph{jets}. For them, DELPHI has an extra particle ID detector
known as RICH. However, to detect heavy hadrons in nature, it is
important to remind that they decay weakly, sometimes in leptons
with long lifetime and characteristic masses and event shapes. For instance, $b$ and $c$ hadrons decay in to the leptons  around $20\%$ with high momentum $p$. The electrons are then ionized in tracking chambers while the muons match between the central track and muon chambers. Moreover, the leptons give charge to the decaying hadron as in $e^+e^-\rightarrow Z\rightarrow b\bar{b}$ in L3. However, in the LHC
a Higgs mass of up to twice of the $Z$ boson mass might be measured.
The production mode is based on partonic processes, as in the
Tevatron, and the greatest rate should come from gluon fusion to
form a Higgs particle ($gg\rightarrow H$) via an intermediate
top-quark loop where the gluons produce a virtual top-quark pair
which couples to the Higgs particles. Furthermore, the alternatives
are the channels of hadronic jets, with a richer kinematic structure
of the events, which should allow refined cuts increasing the
signal-to-background ratio. The latter channels are the quark-gluon
scattering ($q(\bar{q})g\rightarrow q\bar{q}H$) and the
quark-antiquark annihilation ($q\bar{q}\rightarrow gH$), both
dominated by loop-induced processes involving effective $ggH$ and
$ggHZ$-couplings \cite{Bre04}. Nevertheless, there is still the
possibility of more decaying channels, and the generalizations of SM
(for example in supersymmetric models) demands the existence of more
possible decays with supersymmetric particles (such as through
squark loops). Such a generalization might be needed with respect to
the problems those are not solvable within SM and seem to be
definitely secured within the supersymmetric version of SM (i.e.
MSSM). However, experimental evidences for supersymmetric particles
are important to sustain the physical reality of the theory as well
as to clarify the reason of such heavy masses (if these particles
really do exist). In fact, the self-consistency of SM to GUT at a
scale of about $10^{16}$GeV requires a Higgs mass with the upper and
lower bounds (2003) as given below \cite{Wel03},
\begin{align}
130\, \text{GeV}\lesssim M_H\lesssim 190\,\text{GeV},
\end{align}
and which was corrected in 2004 (after exact measurements of top
quark mass) with the following value,
\begin{align}
M_H\lesssim 250\,\text{GeV}\label{bound}.
\end{align}
Such higher values of mass for the Higgs bosons make the theory
non-perturbative while too low values make vacuum unstable
\cite{Wel03}. From the experimental point of view and within the
standard models, the Higgs masses $M_H < 114$GeV are excluded at
least with 95 \%  of the what we call confidence level. The Fermi
laboratory also announced an estimate of 117GeV for the same in
June 2004. However, the EW data strongly prefers the light Higgs
bosons and according to the fit precision of all data, the most
likely value of Higgs mass should be slightly below the limit set
by the direct searches at LEP2, while the upper limit for Higgs
mass lies around $220$GeV at 95\% of confidence level
\cite{D0Col}, which is given by equation (\ref{bound}) after
making all the relevant corrections. Moreover, it would be quite
interesting to look for an existing theoretical prediction of a
Higgs mass around 170GeV \cite{Cha06}, which comes
from an effective unified theory of SM based on noncommutative geometry and with neutrino-mixing coupled to gravity. The Higgs boson mass is,
however, basically unrestricted, but even there are even
indications from lattice calculations that the simplest version of
EW interaction is inconsistent unless $ M_H \lesssim 700$GeV
\cite{Wel03}. With such a mass higher than $800$GeV, the Higgs
bosons would be strongly interacting, so that many new signals
would appear in the Higgs boson scenario. There are also some  evidences for the indirect searches of Higgs scalar in Ultra High Energy (UHE) cosmic rays interactions \cite{udg}. {\it Where the Higgs particles do appear} ? is of course an open question, but an important puzzle to solve in the elementary particle physics, and it may be consisting of some new conjectures which have still to come in modern physics.
\section{Concluding remarks}
\noindent The SM of modern elementary particle physics provides a
concise and accurate description of all fundamental interactions
except gravitation. The answer of the fundamental problem which
allows the elementary particles to become heavy is now addressed in
terms of the Higgs boson in SM, which is quite unlike either a matter
or a force particle. The Higgs Mechanism is, therefore, a powerful
tool of modern particle physics which makes the models
mathematically consistent and able to explain the nature of
fundamental interactions in a manifest way. The bosons and fermions
are believed to gain mass through a phase transition via Higgs
Mechanism. In this way, the particles are able to be coupled with
experiments, and a theoretical explanation may be given of how the
mass generation takes place. Nevertheless, the Higgs particles,
belonging to the Higgs field, are still not experimental reality and
need to be observed to make any model complete. In the
same way, the SM might be needed to be generalized consistently in
view of the different hues of unification schemes and other models,
viz GUT, SuSy, TC, STT (with or without Higgs Mechanism). One more
possibility to answer the problem comes from the LQG, which seems to
explain the general nature of all particles in space-time.\\
The Higgs particles interact in a gravitative and Yukawa form, but
their nature is still not completely understood. Their
fundamental existence is not a fact until and unless they are observed
in high energy experiments such that the SSB could finally be believed to be the natural process of the mass generation mechanism. On the other hand, the Higgs particles may turn out to couple only
gravitationally with a possibility to be generated in the
accelerators in form of the SIDM. The search for Higgs particles is a very important task in physics and it is believed that their mass
would be achievable with the future generation of high energy
experiments (especially in those which are scheduled to
start at LHC in near future). In a lucid way of speaking, the Higgs bosons are believed to have a mass of less than $250$GeV and over $130$GeV, and the current experimental status is that they are heavier than $114$GeV. The search for the Higgs boson is still a matter of speculation in the absence of clear experimental evidences, and the detection of Higgs particles as a real observable particle in future will be a momentous occasion ({\it eureka moment}) in
the world of elementary particle physics to certify the basic ideas
of SSB for the mass generation in the universe. \\ \\  {\bf
Acknowledgments} : The authors are grateful to Professor H. Dehnen for
his kind motivation and help to substantially improve the
manuscript and also thankful for his warm hospitality during their stay
at the Fachbereich Physik, Universit$\ddot{a}$t Konstanz, Germany.
The authors would like to sincerely thank Professor F. Steiner and R.
Lamon, Universit$\ddot{a}$t Ulm, Germany for various interesting as well as stimulating discussions and useful comments.
\newpage
\bibliography{apssamp}

\begin{thebibliography}{99}
\small{
\bibitem{Weinberg} S. Weinberg, \textit{Gravitation and Cosmology:
Principles and Applications of General Theory of Relativity}
 (New York: John Wiley \& Sons Inc.) (2004)
\bibitem{Hartlee} J. B. Hartlee, \textit{Gravity: An Introduction to
Einstein's General Relativity} (Singapore: Pearson Education Inc)
(2003)
\bibitem{Kan93} G. L. Kane, \textit{Perspectives on Higgs Physics I} (Singapore : World Scientific) (1993) and references therein
\bibitem{Peskin} M. E. Peskin and D. V. Schroeder, \textit{An Introduction to Quantum Field Throry}  (Massachusetts: Addison Wesley Reading) (1995) and references therein
\bibitem{Vel77} M. Veltman, {\it Acta Phys. Pol.} {\bf B8}
475 (1977).
\bibitem{Gui91} M. Guildry, \textit{Gauge Field Theories: An Introduction
with Applications} (New York: John Wiley \& Sons, Inc.) (1991)
\bibitem{Wet05} C. Wetterich, {\it Phys. Rev. Lett.} {\bf 94} 01160 (2005)
\bibitem{Gre95} W. Greiner and B. M$\ddot u$ller, \textit{Eichtheorie der schwachen Wechselwirkung} (Frankfurt: Harri Deutsch) (1995)
\bibitem{Hil87} A. Hill and J. J. van der Bij, {\it Phys. Rev.} {\bf D16} 3463 (1987)
\bibitem{Bij95} J. J. van der Bij, {\it Int. J. Theor. Phys.} {\bf 1} 63 (1995);
hep-ph/9507389 v1 (Based on the talks given at the DPG meeting,
Dortmund, 1-4 March, 1994 and Bad Honnef, 7-10 march, 1994)
\bibitem{Deh91} H. Dehnen  and H. Frommert, {\it Int. J. Theor. Phys.}
{\bf 30} 985 (1991)
\bibitem{Bij94} J. J. van der Bij, {\it Acta Phys. Pol.} {\bf B25}
827 (1994)
\bibitem{Bez07a} N. M. Bezares-Roder and H. Dehnen, {\it Gen. Relativ.
Gravit.} {\bf 39} 1259 (2007)
\bibitem{Mand} S. Mandelstam, {\it Phys. Rep.} {\bf 23C} (1976) 245\\
H. Nandan, H. C. Chandola and H. Dehnen, {\it Int. J. Theor. Phys.} {\bf 44} (2005) 469
\bibitem{Hn1} H. Nandan, T. Anna and H. C. Chandola,  {\it Euro
Phys. Letts.} {\bf 67} (2004) 746  and references therein\\
H Nandan {\it AIP Conference Proceedings}
{\bf 393} 16 (2007)
\bibitem{Hig64} P. W. Higgs, {\it Phys. Rev. Lett.}  {\bf 12} 132 (1964); ibid {\bf 13} 509 (1964)
\bibitem{Deh95} H.  Dehnen and E. Hitzer, {\it Int. J. Theor. Phys.}
{\bf 34} 1981 (1995)
\bibitem{Gei97} A. Geitner, D. Ketterer and H. Dehnen, {\it Nuov.
Cim.} {\bf B115}(12) 1357 (2000);  gr-qc/9712074
\bibitem{Ran06} A. Randono, gr-qc/0611073; gr-qc/0611074
\bibitem{Moh86} R. N. Mohapatra, \textit{Unification and
Supersymmetry} (Harrisonburg: Springer Verlag) (1986)
\bibitem{Cer96} J. L. Cervantes Cota, \textit{Induced Gravity and
Cosmology} (Konstanz: Hartung-Gorre Verlag) (1996)
\bibitem{Pec77} R. D. Peccei and H. Quinn, {\it Phys. Rev. Lett.}
{\bf 38} 1440 (1977); \emph{ibid} {\bf D16} 1791 (1977)
\bibitem{Wei78} S. Weinberg, {\it Phys. Rev. Lett.} {\bf 40}, 223 (1978)
\bibitem{Zav06} E. Zavattini \emph{et al.}, {\it Phys. Rev. Lett.}
{\bf 96} 110406 (2006)
\bibitem{Kan97} G. L. Kane, \textit{Perspectives on
Higgs Physics II} (Singapore: World Scientific) (1997) and
references therein
\bibitem{Gie03} H. Gies, J. Jaeckel and C. Wetterich, {\it Phys. Rev.}
{\bf D69} 105008 (2004)
\bibitem{Bil06} S. O. Bilson-Thomson, F. Markopoulou and L. Smolin,
hep-ph/0603022
\bibitem{Fae-Deh05} A. F$\ddot{a}$ssler and, C. J$\ddot{o}$nsson, \textit{Die Top Ten der sch$\ddot{o}$nsten physikalischen Experimenten} (Hamburg: Rororo Science) (2005)
\bibitem{Deh90} H. Dehnen and H. Frommert, {\it Int. J. Theor. Phys.}
{\bf 29} 361 (1990)
\bibitem{Fau01} B. Fauser, {\it Gen. Relativ. Gravit.} {\bf 33} 875 (2001)
\bibitem{Jor55} P. Jordan, \textit{Schwerkraft und Weltall} (Braunschweig : Vieweg \& Sohn Verlag) (1955)
\bibitem{Cot97} S. Cotsakis, gr-qc/9712046 (Talk presented at the 8th
Marcel Grossmann Meeting at Jerousalem) June 22-27 (1997)
\bibitem{Bra61} C. Brans and R. Dicke, {\it Phys. Rev.}
{\bf 124} 925 (1961)
\bibitem{OHan72} J. O'Hanlon, {\it Phys. Rev. Lett.}  {\bf 29} 137 (1972)
\bibitem{Ach73} R. Acharya and P. A. Hogan, {\it Nuov. Cim.
Lett.} {\bf 6} 668 (1973)
\bibitem{Fuj74} Y. Fujii, {\it Phys. Rev.} {\bf D9} 874 (1974)
\bibitem{Zee79} A. Zee, {\it Phys. Rev. Lett.} {\bf 42} 417 (1979)
\bibitem{Berg68} P. G. Bergmann, {\it Int. J. Theor. Phys.} {\bf 1} 25 (1968)
\bibitem{Bro01} K. A. Bronnikov, {\it Acta Phys. Pol.} {\bf B32} 3571
(2001)
\bibitem{Ein13} A. Einstein, {\it Phys. Zs.} {\bf 14} 1260 (1913)
\bibitem{Deh92} H. Dehnen, H. Frommert and F. Ghaboussi,
{\it Int. J. Theor. Phys.}  {\bf 31} 109 (1992)
\bibitem{Deh93} H. Dehnen and H. Frommert, {\it Int. J. Theor. Phys.}
{\bf 32} 1135  (1993) and references therein
\bibitem{Ges92} E. Gessner, {\it Astrophys. Sp. Sc.}
{\bf 196} 29 (1992)
\bibitem{San86} R. H. Sanders, {\it Astron. Astrophys.}
{\bf 154} 135 (1986)
\bibitem{Mil83} M. Milgrom, {\it Astrophys. J.} {\bf 279} 370 (1983)
\bibitem{Ben00}  M. C. Bento, O. Bertolami, R. Rosenfeld and
L. Teodoro, {\it Phys. Rev.} {\bf D62} 041302 (2000)
\bibitem{Boe03} C. B{\oe}m  and P. Fayet, {\it Nucl. Phys.}
{\bf B683} 219 (2004)
\bibitem{Cer95}J. L.  Cervantes-Cota  and H. Dehnen, {\it Phys. Rev.}
{\bf D51} 395 (1995)
\bibitem{Cer95b} J. L. Cervantes-Cota and H. Dehnen,
{\it Nucl. Phys.} {\bf B442} 391 (1995)
\bibitem{Bez07b} N. M. Bezares-Roder, H. Nandan and H. Dehnen,
{\it Int. J. Theor. Phys.} {\bf 46} 2429 (2007)
\bibitem{Per98} S. J. Perlmutter \emph{et al.}, {\it Nature}
{\bf 391} 569 (1998)
\bibitem{Rie98} A. G. Riess \emph{et al.}, {\it Astron. J.}
{\bf 116} 1009 (1998)
\bibitem{Sch79} J. Scherk, {\it Phys. Lett.} {\bf B88} 265 (1979)
\bibitem{Gut81}A. Guth, {\it Phys. Rev.} {\bf D23} 347 (1981)
\bibitem{Lin05} A. Linde, {\it J. Phys. Conf. Ser.} {\bf 24} 151 (2005)
\bibitem{Pen65} R. Penrose, {\it Phys. Rev. Lett.} {\bf 14} 57 (1965)
\bibitem{Ger66} R. Geroch, {\it Phys. Rev. Lett.} {\bf 17} 445 (1966)
\bibitem{Haw68} S. Hawking and G. Ellis, {\it Astrophys. J.}
{\bf 152} 25 (1986)
\bibitem{Lib69} L. M. Libby and F. J. Thomas, {\it Phys. Lett.}
{\bf B30} 88 (1969)
\bibitem{Deh75} H. Dehnen and H. H$\ddot{o}$nl, {\it Astrophys. Sp. Sc.} {\bf 33} 49 (1975)
\bibitem{Ash06}A. Ashtekar, T. Pawlowski and P. Singh, {\it Phys. Rev. Lett.} {\bf 96} 141301 (2006)
\bibitem{Lee07} J. S. Lee and S. Scopel, hep-ph/070122
\bibitem{Wel03} P. Wells, The LEP Saga:  $http://agenda.cern.ch
(2003)$
\bibitem{Bre04} O. Brein and W. Hollik, {\it Phys. Rev.} {\bf D68} 095006 (2003); hep-ph/0402058
\bibitem{D0Col} Y. P. Viala \emph{et al.} (D0 Collaboration), {\it Nature}
{\bf 429} 638 (2004)
\bibitem{Cha06} A. H. Chamseddine, A. Connes and M. Marcolli, hep-th/0610241
\bibitem {udg} U. D. Goswami and K. Boruah, \textit{Czechoslovak Journal of Physics} {\bf 55} 657 (2005) and references therein}
\end{thebibliography}

\end{document}